\documentclass[fleqn,usenatbib]{mn2e}

\usepackage[letterpaper, top = 1.in, bottom=1.in, left=0.95in, right=0.95in]{geometry}

\usepackage[usenames,dvipsnames]{color}
\usepackage{epsfig}
\usepackage{amsmath}
\usepackage{amssymb}
\usepackage[hyphens]{url}

\newcommand{\Msun}{M_{\odot}}

\begin{document}

\title[CFHTLenS Cluster Shear]{CFHTLenS: a weak lensing shear analysis of the 3D-Matched-Filter galaxy clusters}

\author[J. Ford et al.]
{Jes Ford,$^{1}$\thanks{Email: jesford@phas.ubc.ca}
Ludovic Van Waerbeke,$^1$
Martha Milkeraitis,$^1$
Clotilde Laigle,$^{1,2,3}$ \newauthor
Hendrik Hildebrandt,$^{1,4}$
Thomas Erben,$^4$ 
Catherine Heymans,$^5$
Henk Hoekstra,$^6$ \newauthor
Thomas Kitching,$^7$
Yannick Mellier,$^{2,8}$
Lance Miller,$^9$
Ami Choi,$^5$ \newauthor
Jean Coupon,$^{10,11}$
Liping Fu,$^{12}$
Michael J. Hudson,$^{13,14}$
Konrad Kuijken,$^6$ \newauthor
Naomi Robertson,$^5$
Barnaby Rowe,$^{15,16}$
Tim Schrabback,$^{4,6,17}$
Malin Velander$^{6,9}$ \\ \\
$^1$ Department of Physics and Astronomy, University of British Columbia, 6224 Agricultural Rd, Vancouver BC, V6T 1Z1, Canada \\
$^2$ Institut d\'Astrophysique de Paris, UMR7095 CNRS, Universit\'e Pierre \& Marie Curie, 98 bis boulevard Arago, F-75014 Paris, France \\
$^3$ Ecole Polytechnique, 91128 Palaiseau Cedex, France \\
$^4$ Argelander-Institut f\"ur Astronomie, Auf dem H\"ugel 71, 53121 Bonn, Germany \\
$^5$ The Scottish Universities Physics Alliance, Institute for Astronomy, University of Edinburgh, Blackford Hill, Edinburgh EH9 3HJ, UK \\ 
$^6$ Leiden Observatory, Leiden University, Niels Bohrweg 2, NL-2333 CA Leiden, the Netherlands \\ 
$^7$ Mullard Space Science Laboratory, University College London, Holmbury St Mary, Dorking, Surrey RH5 6NT, UK \\
$^8$ CEA/Irfu/SAp Saclay, Laboratoire AIM, F-91191 Gif-sur-Yvette, France \\
$^9$ Department of Physics, Oxford University, Keble Road, Oxford OX1 3RH, UK \\
$^{10}$ Astronomical Observatory of the University of Geneva, ch. dEcogia 16, 1290 Versoix, Switzerland \\
$^{11}$ Institute of Astronomy and Astrophysics, Academia Sinica, PO Box 23-141, Taipei 10617, Taiwan \\
$^{12}$ Shanghai Key Lab for Astrophysics, Shanghai Normal University, 100 Guilin Road, 200234, Shanghai, China \\
$^{13}$ Department of Physics \& Astronomy, University of Waterloo, Waterloo, ON, N2L 3G1, Canada \\
$^{14}$ Perimeter Institute for Theoretical Physics, 31 Caroline St. N., Waterloo, ON, N2L 2Y5, Canada \\
$^{15}$ Department of Physics and Astronomy, University College London, Gower Street, London WC1E 6BT, UK \\
$^{16}$ Jet Propulsion Laboratory, California Institute of Technology, MS 300315, 4800 Oak Grove Drive, Pasadena CA 91109, USA \\
$^{17}$ Kavli Institute for Particle Astrophysics and Cosmology, Stanford University, 382 Via Pueblo Mall, Stanford, CA 94305-4060, USA
}

\maketitle

\begin{abstract} 
We present the cluster mass-richness scaling relation calibrated by a weak lensing analysis of $\ga$18000 galaxy cluster candidates in the Canada-France-Hawaii Telescope Lensing Survey (CFHTLenS). Detected using the 3D-Matched-Filter cluster-finder of Milkeraitis et al., these cluster candidates span a wide range of masses, from the small group scale up to $\sim10^{15} M_{\odot}$, and redshifts 0.2 $\lesssim z\lesssim$ 0.9. The total significance of the stacked shear measurement amounts to 54$\sigma$. We compare cluster masses determined using weak lensing shear and magnification, finding the measurements in individual richness bins to yield 1$\sigma$ compatibility, but with magnification estimates biased low. This first direct mass comparison yields important insights for improving the systematics handling of future lensing magnification work. In addition, we confirm analyses that suggest cluster miscentring has an important effect on the observed 3D-MF halo profiles, and we quantify this by fitting for projected cluster centroid offsets, which are typically $\sim$ 0.4 arcmin. We bin the cluster candidates as a function of redshift, finding similar cluster masses and richness across the full range up to $z \sim$ 0.9. We measure the 3D-MF mass-richness scaling relation $M_{200 } = M_0 (N_{200} / 20)^\beta$. We find a normalization $M_0 \sim (2.7^{+0.5}_{-0.4}) \times 10^{13} M_{\odot}$, and a logarithmic slope of $\beta \sim 1.4 \pm 0.1$, both of which are in 1$\sigma$ agreement with results from the magnification analysis. We find no evidence for a redshift-dependence of the normalization. The CFHTLenS 3D-MF cluster catalogue is now available at \url{cfhtlens.org}.
\end{abstract}



\section{Introduction}
\label{intro}
The evolution of large scale structure is overwhelmingly driven by the invisible components which make up the majority of the present day energy density of the Universe. In order to probe these structures we are forced to rely on biased tracers of the underlying density field that we can actually observe, such as galaxies. Large galaxy cluster surveys are invaluable in providing sufficient statistics for classifying and analysing the most massive gravitationally bound systems that have had time to form in our cosmic history. In addition to providing a cosmological probe, they are interesting laboratories for the evolution of individual galaxies and the intracluster medium \citep{Voit05}.

Several methods have been developed for identifying clusters in optical galaxy surveys, including the red sequence technique \citep{Gladders00}, density maps \citep{Adami10}, redMaPPer \citep{Rykoff14}, and matched-filter methods \citep{Postman96}. An extension of the latter, 3D-Matched-Filter (3D-MF), is described in \citet{Milkeraitis10} and used in this work. This cluster finder attempts to circumvent the common issue of line-of-sight projections by using photometric redshift information to identify clusters in redshift slices. Beyond the use of photometric redshifts, 3D-MF does not apply any additional colour-selection criteria for identifying clusters (e.g. that cluster members must fall on the red sequence). A similar algorithm tuned for galaxy groups was introduced by \citet{Gillis11}. Every cluster-finding technique will pick out clusters with somewhat distinct characteristics because of different assumptions that are made in the algorithm, and it is therefore important to characterize and contrast independent samples of clusters \citep{Milkeraitis10}.

Among the broad array of analysis tools employed by the galaxy cluster research community, gravitational lensing is a crucial technique for obtaining masses and density profiles, independent of assumptions regarding cluster dynamical state. In the weak regime, lensing provides robust measurements of stacked cluster samples (and individual masses for very massive clusters), affording a statistical view of average galaxy cluster properties \citep{Hoekstra13}. The majority of weak lensing studies measure the shear, or shape distortion, of lensed source galaxies. The complementary magnification component of the lensing signal has more recently been measured with increasing precision \citep{Scranton05,Hildebrandt09b,Ford12,Ford14,Morrison12,Hildebrandt13,Bauer14}, and has been combined with shear in joint-lensing analyses \citep{Umetsu11,Umetsu14}. When combined with other cluster observables, lensing yields useful scaling relations that can be extrapolated with some caution to wider cluster populations, or cross-examined to characterize intrinsic disparities that may distinguish catalogues compiled using different cluster-finding techniques \citep{Hoekstra07,Johnston07,Leauthaud10,Hoekstra12,Covone14,Oguri14}.

Section 2 of this paper describes the data, Section 3 gives the formalism of the weak lensing measurement, and Section 4 presents the results. We then discuss and compare our findings to other results, including our previous magnification measurements of the same lens sample, in Section 5. We finish with conclusions in Section 6. Throughout this work we use a concordance $\Lambda$ cold dark matter cosmology with $\Omega_M$ = 0.3, $\Omega_{\Lambda}$ = 0.7, and H$_0$ = 70 km/s/Mpc.


\section{Data}
\label{data}


\subsection{The Canada-France-Hawaii Telescope Legacy Survey Wide}

The Canada-France-Hawaii Telescope Legacy Survey (CFHTLS) is a multi-component optical survey conducted over more than 2300 h in 5 yr ($\sim450$ nights) using the wide field optical imaging camera MegaCam on the CFHT's imaging system MegaPrime. The Wide survey is composed of four patches ranging from 25-72 deg$^2$, together totalling an effective survey area of $\sim$ 154 deg$^2$. The data were acquired through five filters: $u$*, $g'$, $r'$, $i'$, $z'$, and has a 5$\sigma$ point source $i'-$band limiting magnitude of 24.5. The breadth of CFHTLS-Wide was intended for the study of large scale structure and matter distribution in the Universe.

The CFHTLS-Wide optical multi-colour catalogues used in this work were created from stacked images of the aforementioned Wide fields \citep[see][for details on the data processing and multi-colour catalogue creation]{Erben09, Hildebrandt09a, Hildebrandt12, Erben13}. Basic photometric redshift ($z_{\mathrm{phot}}$) statistics were determined by \citet{Hildebrandt12}. In this work we restrict ourselves to a redshift range of 0.1 $\leq z \leq$ 1.2, which has outlier rates $\lesssim$ 6\% and scatter $\sigma \lesssim$ 0.06.


\subsection{CFHTLenS Shear Catalogue}

The Canada-France-Hawaii Telescope Lensing Survey (CFHTLenS) reduced CFHTLS-Wide data for weak lensing science applications \citep{Heymans12,Erben13}. Many factors affect high-precision weak lensing analyses, including correlated background noise, PSF measurement, and galaxy morphology evolution for example \citep[for a more detailed list and study, see][]{step2,Heymans12}. The efforts of CFHTLenS have led to new reduction methodologies with reduced systematic errors and a more thorough understanding of the PSF and its variation in the CFHTLS-Wide images. As part of this pipeline, {\em lens}fit was used to measure galaxy shapes \citep{Miller13}, which were tested for systematics in \citet{Heymans12}. The galaxy shear measurements and photometric redshifts used in this work are publicly available.\footnote[1]{\url{www.cfhtlens.org}; Data products are made available at \url{http://www.cadc-ccda.hia-iha.nrc-cnrc.gc.ca/community/CFHTLens/query.html}}


\subsection{3D-MF Clusters}\label{3DMF}

Here we give a brief overview of the 3D-MF galaxy cluster-finding algorithm. For additional background and details on the algorithm, including extensive testing on the Millennium Simulation data set, and information on the completeness and purity of a 3D-MF derived galaxy cluster catalogue, the reader is directed to \citet{Milkeraitis10}.

3D-MF searches survey data for areas that maximally match a given luminosity and radial profile for a fiducial galaxy cluster, similar to the technique used by \citet{Postman96}.  For the luminosity profile we use an integrable Schechter function, given by
\begin{equation}
\Phi(M)=0.4 \ \ln(10) \ \Phi^* 10^{0.4(\alpha + 1)(M^*-M)} \exp \left[ -10^{0.4(M^*-M)} \right],
\end{equation}
where $\Phi$ is the galaxy luminosity function, $\Phi^*$ sets the overall normalization, $M$ is absolute magnitude, $M^*$ is a characteristic absolute magnitude, and $\alpha$ is the faint end slope of the luminosity function. As discussed in \citet{Milkeraitis10}, the multiplicative term, exp$[-10^{0.4(M^{*}-M)}]$, keeps this function from diverging when $\alpha < -1$ and $M < M^*$. For the radial profile we use a truncated Hubble profile, given by
\begin{equation}
P \left ( \frac{r}{r_c} \right )=\frac{1}{\sqrt{1+ \left ( \frac{r}{r_c} \right ) ^2}} - \frac{1}{\sqrt{1+ \left ( \frac{r_{co}}{r_c} \right ) ^2}},
\end{equation}
where $r_c$ is the cluster core radius, and $r_{co} \gg r_c$ is the cutoff radius. In an attempt to match both of the above profiles, 3D-MF creates likelihood maps of the sky survey area. Peaks in this map are possible cluster detections, and are each assigned a significance $\sigma_{\rm cl}$ relative to the background signal \citep[$\sigma_{\rm cl}$ is calculated using Equation 5 of][which the reader is referred to for more details]{Milkeraitis10}. The cluster centres are defined to be the locations of the likelihood peaks; see Section \ref{sec:miscentring} for how uncertainties in the centres are dealt with.

An important characteristic of this cluster-finding algorithm is the fact that the described process is carried out in discrete redshift bins to avoid spurious false-detections due to line-of-sight projections. 3D-MF was run on the CFHTLS-Wide catalogues with redshift slices of width $\Delta z = 0.2$, which are then shifted by 0.1, and the finder is run again on the overlapping redshift slices. Clusters are assigned a final redshift estimate (of bin width $\Delta z = 0.1$) by using the centre of the slice that maximizes cluster detection significance. 3D-MF was run using the same run-time parameters listed in table~2 in \citet{Milkeraitis10}, with the exception of an absolute $i'-$band magnitude of $M^*_{i'-\mathrm{band}}=-23.22 \pm 0.01$ and slope of the Schechter luminosity function, $\alpha=-1.04 \pm 0.01$, derived from the Wide data \citep{MMthesis11}.

Excluding possible multiple detections, a total of 22,694 galaxy cluster candidates were found in the CFHTLS-Wide data set with detection significance $\sigma_{\rm cl} \ge 3.5$. Using 3D-MF's multiple detection criteria, there were $34.4\%$ additional duplicate detections of galaxy clusters. This is comparable to the $\sim 36\%$ multiple detection rate found from Millennium Simulation tests and $37.6\%$ found in the CFHTLS-Deep galaxy cluster catalogue in \citet{Milkeraitis10}. Using the Millennium Simulation, \citet{Milkeraitis10} determined that there are potentially $\sim 16\%-24\%$ false positives in 3D-MF-derived galaxy cluster catalogues, distributed mostly in the lower significance ranges \citep[see Table~3 in][]{Milkeraitis10}.

Following the 3D-MF methodology for galaxy cluster catalogue generation, the significance of galaxy cluster detections was used to select the best galaxy cluster candidate among multiple detections, and the remaining multiple detections were rejected from the analysis. A single detection of each cluster candidate then makes up the CFHTLS-Wide galaxy cluster candidate catalogue. We restrict our analysis herein to a cluster redshift range of 0.2 $\lesssim z \lesssim$ 0.9, where 3D-MF detections are the most reliable.

\begin{figure}
\vspace{0.2cm}
  \includegraphics[scale=0.4]{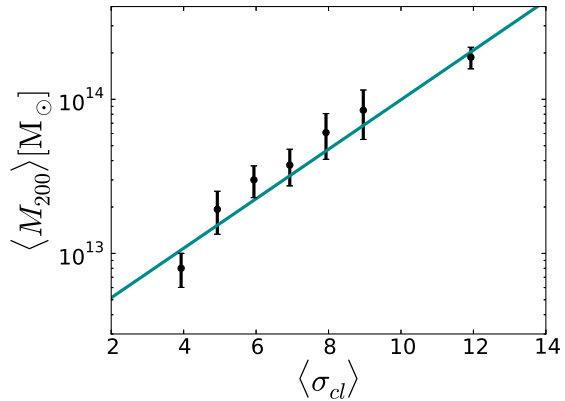}
  \caption{Scaling of shear-measured mass $M_{200}$ with the 3D-MF cluster detection significance $\sigma_{\rm cl}$. Since we find significance to be a good proxy for mass, we use the derived mass-significance relation to estimate a radius $r_{200}$ for each cluster candidate, within which we count galaxies for richness $N_{200}$, as described in the Section \ref{3DMF}.}
\label{plot:masssig} 
\end{figure}

\begin{figure*}
\vspace{0.5cm}
  \includegraphics[width=0.7\textwidth]{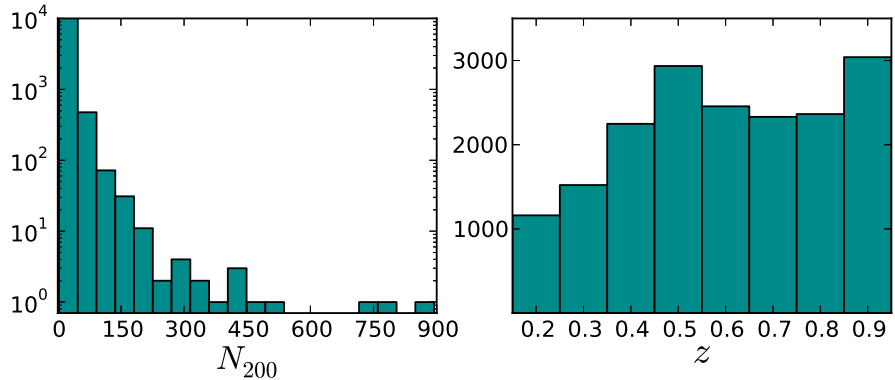}
  \caption{Number of 3D-MF cluster candidates as a function of richness $N_{200}$ and redshift $z$.}
\label{plot:hists}
\end{figure*}

In \citet{Ford14}, we described our method of calculating richness for each of these candidate clusters. $N_{200}$ is defined to be the number of member galaxies brighter than absolute magnitude $M_i \ge -19.35$, which is chosen to match the limiting magnitude at the furthest cluster redshift that we probe ($N_{200}$ is background-subtracted; there is no correction for passive evolution). To be considered a cluster member, a galaxy must lie within a projected radius $R_{200}$ of a cluster centre, and have $\Delta z < 0.08(1+z)$ \citep[based on the photometric errors of the CFHTLenS catalogue; for details regarding $N_{200}$ see][]{Ford14}. $R_{200}$ is defined as radius within which the average density is 200 times the critical energy density of the Universe ($M_{200}$ is the total mass inside $R_{200}$), and in this work has been re-estimated from the data as follows. 

Initially cluster candidates were stacked in bins of cluster detection significance $\sigma_{cl}$, which was found to correlate well with the amplitude of the measured shear profiles, and therefore with mass (see Figure \ref{plot:masssig}). These preliminary masses were estimated using the same method described in Section \ref{method}. A new mass-significance relationship,
\begin{equation}
\label{eqn:masssig}
\mathrm{log}\left[ \frac{M_{200}^{\rm prelim}}{M_{\odot}} \right] = \left( 0.161^{+0.006}_{-0.009} \right) \sigma_{\rm cl} + 12.39^{+0.05}_{-0.08},
\end{equation}
was derived from this result and the preliminary mass values converted into the corresponding radii, which were used to count galaxies for richness ($\sigma_{\rm cl} \rightarrow M_{200}^{\rm prelim} \rightarrow R_{200} \rightarrow N_{200}$). Compared with the richness estimates used in \citet{Ford14}, which were based on a preliminary shear analysis using a more basic cluster modelling approach \citep{MMthesis11}, the updated richnesses are larger in most cases (see the Full Model description in Section \ref{sec:halomodel} for improvements). For the log-normal curve in Figure \ref{plot:masssig}, as well as for all models fit in this work, the best fit is the curve that minimizes $\chi^2$, using a downhill simplex algorithm to search parameter space.

\begin{figure}
\vspace{0.2cm}
  \includegraphics[scale=0.4]{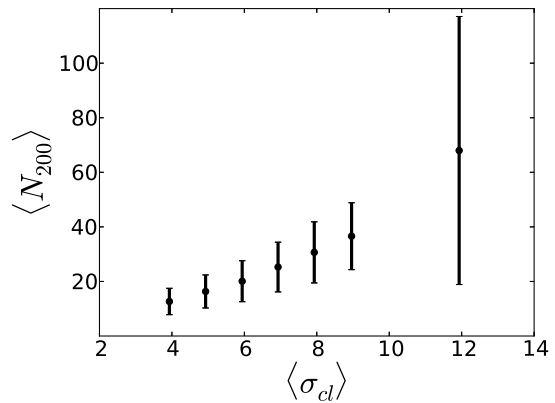}
  \caption{Scaling of richness $N_{200}$ with the 3D-MF cluster detection significance $\sigma_{\rm cl}$. Error bars denote the standard deviation of the ensemble of $N_{200}$ values in each $\sigma_{\rm cl}$ bin. Since $N_{200}$ is estimated using individual cluster radii calculated from the mass-significance relation (Equation \ref{eqn:masssig}), this figure confirms what we would expect -- a strong scaling between richness and significance.}
\label{plot:n200sig} 
\end{figure}

Cluster candidates used in this work are required to have at least $N_{200} > 2$, and a detection significance $\ge 3.5$. The richness and redshift distributions are summarized in Figure \ref{plot:hists}. Figure \ref{plot:n200sig} shows the relative scaling between richness and detection significance. The final catalogue contains the same 18036 cluster candidates used in \citet{Ford14}, now with updated richness estimates based on the shear mass-significance scaling just described. There are also 20 additional low-significance cluster candidates whose revised $N_{200}$ now survive the cuts -- these systems have negligible impact on the overall results, but do increase the total number of clusters to 18056. The full 3D-MF catalogue is available at \url{cfhtlens.org}.


\section{Method}
\label{method}

\subsection{Stacking Galaxy Clusters}

The mass of a galaxy cluster can be determined by measuring shear in binned annuli out from the cluster centre, and fitting this with a theoretical density profile. For the most massive galaxy clusters, this is relatively straightforward. However, for most galaxy clusters (especially given the high number of lower mass galaxy cluster candidates explored in this work), the background noise overwhelms the measurable shear.  Fortunately, stacking many individual galaxy clusters together improves the signal-to-noise ratio, enabling the measurement of a statistically significant signal, averaged over a cluster ensemble.

To obtain a meaningful average for a property of an ensemble of galaxy clusters, similar clusters must clearly be chosen for a stack. It is desirable to stack clusters of very similar mass (and thus clusters of roughly the same size and profile), as an average mass measurement of the cluster stacks is the goal. In fitting models to the stacked weak lensing measurements in this work, we assume that the haloes are spherical on average. However, recent studies have explored halo orientation bias in simulations, demonstrating that optically-selected clusters will tend to be aligned along the line-of-sight, and this effect could lead to our mass estimates being biased high by $3-6$\% \citep{Dietrich14}.

For this analysis, the cluster candidates are stacked in bins of richness $N_{200}$ as well as redshift, identical to those used in \citet{Ford14}. The overall approach is conceptually very similar to that used in galaxy-galaxy lensing \citep[see][]{Velander14}, except we replace the galaxy lenses with cluster lenses.


\subsection{Measuring $\Delta\Sigma$}
\label{sec:measure}

We measure the radial profile of the tangential shear, $\gamma_t(R)$, around each cluster candidate in bins of projected {\it physical} distance $R$, extending from 0.09 to 5 Mpc. The logarithmically-spaced radial bins are chosen to match those used in \citet{Ford14}, which we compare results to in Section \ref{sec:magn}, and the resulting mass measurements are insensitive to small adjustments in the innermost radii. To select background galaxies for measuring shear, we use their redshift probability distributions $P(z_s)$, where $z_s$ is the source redshift. Relative to a given cluster redshift ($z_l$), we require both that (1) the peak of a galaxy's $P(z_s)$ distribution is at higher redshift, and (2) at least 90\% of a galaxy's $P(z_s)$ is at higher redshift. The second requirement is designed to account for the occasional galaxy with an odd $P(z_s)$, which may peak at high redshift (and so would be included in many conventional shear analyses), but could perhaps have a non-negligible tail extending to low $z$, or even be bimodal.

From the individual shear profiles we construct $\Delta\Sigma$, the differential surface mass density, for each stacked cluster candidate sample: 
\begin{equation}
\label{eqn:deltasig}
\Delta\Sigma(R) \equiv \overline{\Sigma}(<R)-\Sigma(R) = \langle \gamma_t(R) \rangle \Sigma_{\mathrm{crit}}.
\end{equation}
Here $\Sigma(R)$ is the surface mass density of a lens, and $\Sigma_{\mathrm{crit}}$ is the critical surface mass density, which depends on the geometry of the lens-source pairs. It is given by
\begin{equation}
\label{eqn:sigcrit}
\Sigma_{\mathrm{crit}} = \frac{C^2}{4 \pi G} \frac{D_s}{D_l D_{ls}},
\end{equation}
where $C$ is the speed of light and $D_s$, $D_l$, and $D_{ls}$, are the angular diameter distances to the source, to the lens, and between the lens and source, respectively. 

In computing $\Sigma_{\mathrm{crit}}$ for {\it each lens-source pair}, we treat the individual lens $z_l$ as fixed, and integrate over the full source $P(z_s)$, for $z_s > z_l$, to compute the distances:
\begin{equation}
D_s = \int_{z_l}^{\infty} D_{ang}(0,z_s) P(z_s) {\rm d}z_s
\end{equation}
\begin{equation}
D_{ls} = \int_{z_l}^{\infty} D_{ang}(z_l,z_s) P(z_s) {\rm d}z_s
\end{equation}
Here $D_{ang}$ is the angular diameter distance between two redshifts (and $D_l$ is simply $D_{ang}(0,z_l)$). The source redshift probability distribution is renormalized behind the lens, so that $\int_{z_l}^{\infty} P(z_s) {\rm d}z_s = 1$. Using the full $P(z_s)$ distribution should improve any residual photo-$z$ calibration bias in the lensing measurement \citep{Mandelbaum08a}.

We follow the same procedure described in detail in \citet{Velander14}, wherein we combine shear profiles using the $lens$fit source weighting \citep[Equation 8 of][]{Miller13}, and apply a correction for multiplicative bias \citep{Miller13}, so that the $\langle \gamma_t(R) \rangle$ appearing in Equation \ref{eqn:deltasig} is the average {\it calibrated} tangential shear. We estimate a covariance matrix for each stacked sample, by running 100 sets of bootstrapped cluster measurements, and calculating the covariance as:
\begin{equation}
\begin{split}
C(R_i,R_j) = \left[ \frac{N}{N-1} \right]^2 \frac{1}{N} \sum_{k=1}^{N} \left[\Delta\Sigma_k(R_i) - \overline{\Delta\Sigma}(R_i)\right] \\
\times \left[\Delta\Sigma_k(R_j) - \overline{\Delta\Sigma}(R_j)\right]
\end{split}
\end{equation}
Here $N$ is the number of bootstrap samples, $R_i$ and $R_j$ denote specific angular bins, and $\overline{\Delta\Sigma}(R_i)$ is the differential surface mass density at $R_i$, averaged across all bootstrap realizations. The square-root of the diagonal of this matrix yields the error bars displayed on the weak lensing measurements in Section \ref{sec:results}. We confirm that $N=100$ bootstrap realizations of the data is sufficient by tracking the covariance estimated from different numbers of bootstrapped samples and checking for convergence, which typically occurs at around 40 realizations. We use the full covariance matrices when fitting to the data, as will be described in Section 4.1.

We test our $\Delta\Sigma$ measurements for systematics by measuring the rotated shear $\gamma_r(R)$ (where each galaxy ellipticity is rotated by 45$^{\circ}$), finding a signal consistent with zero. We also check that masked areas and edge effects are not affecting our measurement, by measuring $\Delta\Sigma$ around many randomly chosen points ($>$ 50 times the number of cluster candidates), and we find no significant signal here either.

\subsubsection{The NFW model}
\label{sec:nfw}

We use the Navarro, Frenk and White (NFW) dark matter density profile \citep{nfw97} for modelling $\Delta\Sigma$. As demonstrated by numerical simulations, the dissipationless collapse of density fluctuations under gravity produces overdensities that are approximated well by the NFW profile
\begin{equation}
\rho_{\mathrm{NFW}}(r) = \frac{\delta_c \rho_{\mathrm{crit}}(z)}{(r/r_s)(1+r/r_s)^2},
\end{equation}
where $\delta_c$ is the characteristic overdensity of a halo, and $\rho_{\mathrm{crit}}(z)$ is the critical energy density of the Universe at that redshift. The scale radius is $r_s = R_{200}/c$, where $c$ is the concentration parameter (not to be confused with the speed of light $C$ in Equation \ref{eqn:sigcrit}). $R_{200}$ is the cluster radius, and the total mass within that radius is known as $M_{200}$. \citet{Wright00} derived the NFW forms of the projected mass density profiles in Equation \ref{eqn:deltasig}, which we make use of in this work.

In general, the NFW profile is a two parameter model for the halo density, commonly parametrized in terms of $M_{200}$ and $c$. However, there is a well-established correlation between these two parameters, and it is common to introduce a mass-concentration relation to reduce the dimensionality of the problem (note that concentration itself may be degenerate with cluster centroid offsets, which will be discussed in Section \ref{sec:miscentring}). In this work we invoke the mass-concentration relation recently presented by \citet{Dutton14} for the Planck cosmological parameters, which successfully characterizes the profiles of simulated haloes spanning a wide range of masses and redshifts. Given a cluster mass, the concentration is then fixed, and we have just a single mass-related fit parameter to deal with.

\begin{figure*}
\vspace{0.8cm}
  \includegraphics[width=0.9\textwidth]{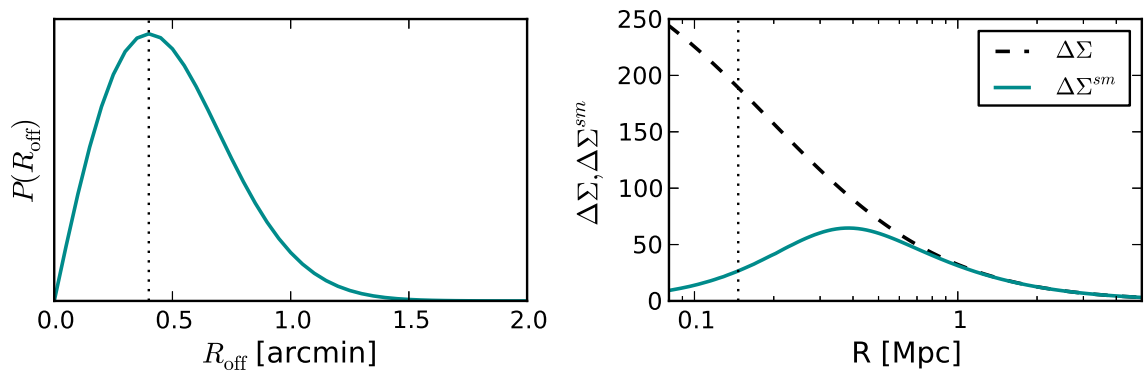}
  \caption{This figure is an illustrative example of typical $\Delta\Sigma(R)$ and $\Delta\Sigma^{\rm sm}(R)$ profiles, to demonstrate the effects of cluster miscentring (Equations 7 -- 11) on measured shear density profiles. The left-hand panel shows a typical probability distribution of centroid offsets, $P(R_{\mathrm{off}})$, modelled via a 2D Gaussian with $\sigma_{\mathrm{off}}$ = 0.4 arcmin. The right-hand panel demonstrates the effect of this offset distribution on the measured shear profile (in vertical axis units of [$\Msun$/pc$^2$]) of a fiducial halo of mass $M_{200}=10^{14}\Msun$, located at $z=0.5$. The dashed black curve shows the perfectly centred $\Delta\Sigma(R)$ profile, and the solid blue curve shows the miscentred profile $\Delta\Sigma^{\rm sm}(R)$. In both panels, the vertical dotted line marks the location of the miscentring offset $\sigma_{\mathrm{off}}$, to guide the eye in the comparison.}
\label{plot:miscentring}
\end{figure*}

\subsubsection{Non-weak shear corrections}

The gravitational lensing observable is galaxy shapes. From these, we measure the reduced shear $g = \gamma / (1-\kappa)$ about the lens, where $\gamma$ is the true shear and $\kappa = \Sigma / \Sigma_{\mathrm{crit}}$ is the convergence \citep[as before, calculated using the NFW halo formalism in][]{Wright00}. At the innermost radii that we probe ($\sim$ 0.1 Mpc) the common weak lensing assumption that $g \approx \gamma$ may break down for the more massive clusters. We account for the difference between true and reduced shear using the correction factor from \citet{Johnston07}, which was worked out in detail in \citet{Mandelbaum06}. The differential surface mass density corrected for non-weak shear is given by:
\begin{equation}
\widehat{\Delta\Sigma} = \Delta\Sigma + \Delta\Sigma \ \Sigma \ \mathcal{L}_z,
\end{equation}
where $\mathcal{L}_z = \langle \Sigma_{\mathrm{crit}}^{-3} \rangle / \langle \Sigma_{\mathrm{crit}}^{-2} \rangle$ is calculated for each cluster redshift, using the full distribution of background galaxies satisfying the same redshift requirements outlined in Section \ref{sec:measure}. Similar to \citet{Leauthaud10}, we ignore any radial variations of $\mathcal{L}_z$, but do account for the variation with redshift, as our cluster sample spans a large $z$ range. The entire correction term $\Delta\Sigma \ \Sigma \ \mathcal{L}_z$ is negligible at all radii except for the innermost bin, where it typically makes up a few percent (at most $\sim$10\%) of the measured signal.


\subsection{Miscentring Formalism}
\label{sec:miscentring}

As was shown in \citet{Milkeraitis10}, 3D-MF does not always determine the exact correct centre for a galaxy cluster, and clusters may not always have a well-defined centre. This is a problem with all galaxy cluster finders and dealing with it properly involves understanding and quantifying its effects, such as including the uncertainty of the centre in calculations. The amplitude of measured shear profiles is absolutely dependent on the declared centre of the profile, so miscentring can potentially have a large impact on results. Offset cluster centres that are mistakenly modelled as being the true centres of the gravitational potentials will lead to underestimates in the inferred lens masses.

In our first analysis of the 3D-MF cluster candidates, we found modest evidence for cluster centroid errors \citep{Ford14}. However, that work relied on the lensing magnification technique, which is less sensitive to these effects than the shear, since magnification directly probes $\Sigma(R)$, while it is $\Delta\Sigma(R)$ that is more drastically reduced by a misplaced centre. See, for example, fig. 4 in \citet{Johnston07}, for a nice illustration of the comparative effect of miscentring on these two lensing profiles. 

In this work, we are able to directly quantify the presence of cluster miscentring by fitting for the offsets in our measurements of $\Delta\Sigma$. As will be shown in Section 4, we find that the best-fitting distribution of centroid offsets is in agreement with the following distribution based on simulations, which we assumed in \citet{Ford14}. 

The distribution of cluster offsets can be modelled as a two-dimensional Gaussian, by using a uniform angular distribution and the following radial profile:
\begin{equation}
P(R_{\mathrm{off}})=\frac{R_{\mathrm{off}}}{\sigma_{\mathrm{off}}^2}\ \mathrm{exp}\bigg[-\frac{1}{2}\bigg(\frac{R_{\mathrm{off}}}{\sigma_{\mathrm{off}}}\bigg)^2\ \bigg].
\end{equation}
Here, $R_{\mathrm{off}}$ is the projected offset of the 3D-MF derived galaxy cluster centre from the true galaxy cluster centre, and $\sigma_{\mathrm{off}}$ is the width of the distribution and one of the miscentring parameters which we fit to the stacked shear measurement. An example $P(R_{\mathrm{off}})$ curve is plotted in the left-hand panel of Figure \ref{plot:miscentring}, for $\sigma_{\mathrm{off}}$=0.4 arcmin. Note that we use physical units (e.g. Mpc) for most distances in this work, the exception being $\sigma_{\mathrm{off}}$ which we report in angular size (arcmin). The reason for this choice is that we believe a significant contribution to miscentring derives from 3D-MF's cluster characterization, which does not for example select a member galaxy as the centre (this choice of angular size is a matter of taste, since complex cluster physics certainly contributes to ambiguous halo centres). 

The effect of this offset distribution $P(R_{\mathrm{off}})$ is to reduce the ideal $\Sigma(R)$ to a smoothed profile \citep[see e.g.][]{Johnston07,George12}
\begin{equation}
\label{EQ:sigsmooth}
\Sigma^{\rm sm}(R)=\int_{0}^{\infty} \Sigma(R|R_{\mathrm{off}})\ P(R_{\mathrm{off}})\ \mathrm{d}R_{\mathrm{off}},
\end{equation}
which is illustrated in the right-hand panel of Figure \ref{plot:miscentring}. Equation \ref{EQ:sigsmooth} is an integration over all possible values of $R_{\mathrm{off}}$ in the distribution. The expression for the surface mass density at a single $R_{\mathrm{off}}$ is
\begin{equation}
\Sigma(R|R_{\mathrm{off}})=\frac{1}{2\pi}\int_{0}^{2\pi}\Sigma(r) \mathrm{d}\theta,
\end{equation}
where $r = \sqrt{R^2+R_{\mathrm{off}}^2-2RR_{\mathrm{off}}\cos(\theta)}$ and $\theta$ is the azimuthal angle \citep{Yang06}. From the smoothed $\Sigma^{\rm sm}(R)$ profile, we can obtain the smoothed shear profile:
\begin{equation}
\Delta\Sigma^{\rm sm} = \overline{\Sigma^{\rm sm}}(<R) - \Sigma^{\rm sm}(R)
\end{equation}
\begin{equation}
\overline{\Sigma^{\rm sm}}(<R) = \frac{2}{R^2} \int_{0}^{R} \Sigma^{\rm sm}(R')R'\mathrm{d}R'
\end{equation}

See \citet{George12} for a discussion of the effects of cluster miscentring on measured shear profiles. There are several different approaches in the literature for actually applying this formalism to data. For example, in some work authors apply the same smoothing to all clusters in a stack \citep{George12}, whereas others apply a two-component smoothing profile \citep{Oguri14}, or chose a uniform distribution of offsets instead of the Gaussian \citep{Sehgal13}. In our previous analysis of this cluster candidate sample, the magnification technique did not give significant constraining power for additional parameters, so we simply compared fits for both a perfectly centred and miscentred model, using estimates of $\sigma_{\mathrm{off}}$ obtained from running 3D-MF on simulations \citep{Ford14}. Both \citet{Johnston07} and \citet{Covone14} applied a combination of perfectly centred and miscentred haloes, thus fitting for the fraction of offset clusters in addition to the magnitude of the offset distribution $\sigma_{\mathrm{off}}$. We follow this latter approach in the current analysis.

As a caveat, we note that the degree of miscentring is fairly degenerate with the cluster concentration parameter, as both can have an effect on the amplitude of the inner shear profile. For example, we tried using the mass-concentration relation of \citet{Prada12}, which yields higher concentration for a given mass than the \citet{Dutton14} relation used here, and results in a best fit with larger centroid offsets. For the lower mass (richness) clusters this change is negligible, but for the most massive clusters in this study, the choice of concentration-mass relation can affect the miscentring fit parameters by as much as 40\%. Importantly, however, the best-fitting cluster mass is the {\it same} in both cases (within the stated 1$\sigma$ uncertainties). The degeneracy of cluster concentration and miscentring would be important to consider in a study seeking to constrain cluster mass-concentration relations. The measured concentrations will be biased low if cluster centroid offsets are significant and not fully accounted for.


\subsection{The Halo Model}
\label{sec:halomodel}
Weak lensing measurements are sensitive to the fact that structures in the Universe are spatially correlated. We account for this large scale clustering using the halo model, which provides a useful framework for modelling the clustered and complex dark matter environments that we probe in gravitational lensing studies. This phenomenological approach places all the matter in the Universe into spherical haloes, which are clustered according to their mass. Observables such as galaxies and clusters are considered biased tracers of the underlying dark matter distribution, with a bias factor that has been constrained in many numerical simulations \citep[e.g.][]{Mo96,Sheth99,Tinker10}. See \citet{Cooray02} for an extensive review of the halo model.

We follow an approach similar to \citet{Johnston07}, in considering a two-halo term in addition to the main NFW halo fit to our weak lensing shear measurement. Calculation of the two-halo term is identical to our approach in \citet{Ford14}, and we refer the reader there for explicit details. The two-halo term is proportional to a cluster bias factor which depends on mass, and for this we continue to use the $b(M)$ relation of \citet{Seljak04}. The full model including the two-halo term is:
\begin{equation}
\Delta\Sigma(R) = p_{\mathrm{cc}}\Delta\Sigma_{\mathrm{NFW}} +(1-p_{\mathrm{cc}})\Delta\Sigma_{\mathrm{NFW}}^{\rm sm} + \Delta\Sigma_{\mathrm{2halo}}
\label{modelEQ}
\end{equation}

The fraction of cluster candidates that is {\it correctly centred} on their parent dark matter haloes, $p_{\mathrm{cc}}$, is a parameter that we fit to the data. $p_{\mathrm{cc}}$ is a continuous variable, bounded between 0 and 1, fit separately for each stacked weak lensing measurement. Thus we have two cluster-centring-related parameters ($p_{\mathrm{cc}}$ and $\sigma_{\mathrm{off}}$), as well as one mass-related parameter ($M_0$), in the final modelling of the data.


\section{Galaxy Cluster Weak Lensing Shear Results}
\label{sec:results}

\begin{figure*}
\vspace{2.cm}
  \includegraphics[width=0.85\textwidth]{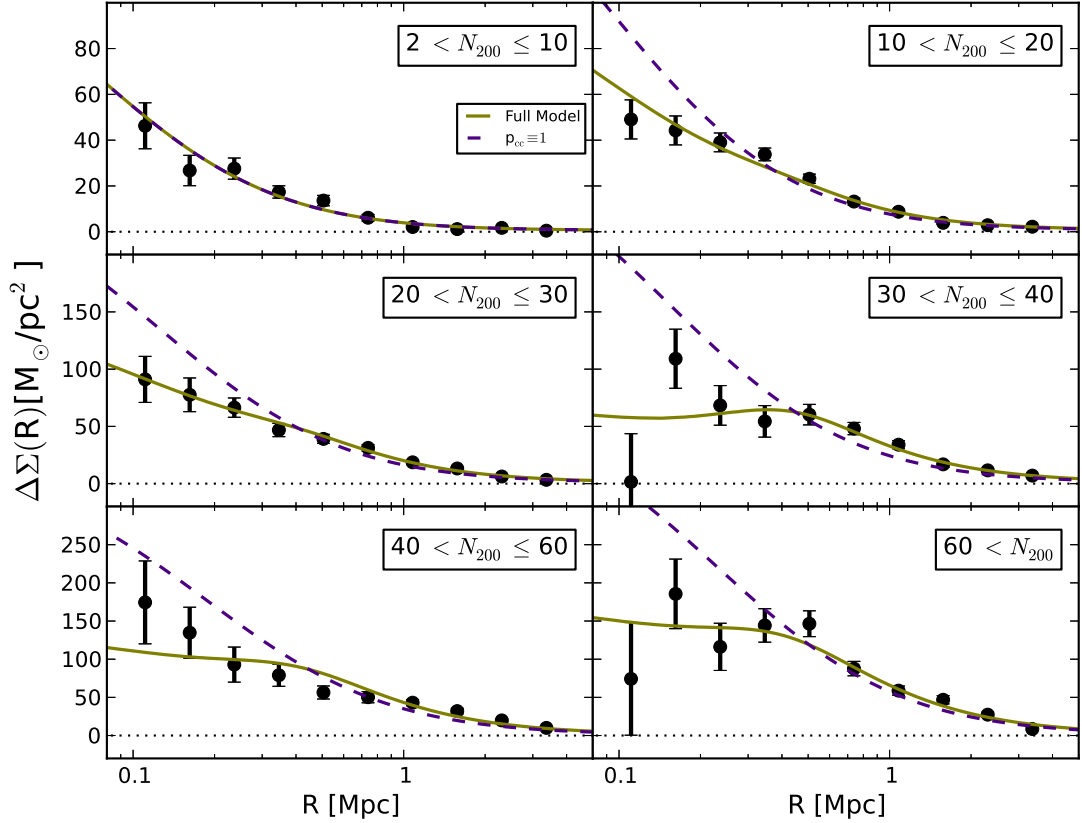}
  \caption{Best-fitting models for each richness-binned stack of cluster candidates. The solid green curves are the best fits to the full model given by Equation \ref{modelEQ}. The dashed purple curves are the best-fitting models which assumes that every cluster centre identified by 3D-MF is perfectly aligned with the dark matter halo centre. With the exception of the lowest richness bin, where the best-fitting curves coincide, the perfectly centred model does not provide a good fit to the data at small $R$. Tables \ref{richbintable1} and \ref{richbintable2} summarize the results of both fits.}
\label{plot:nbinned}
\end{figure*}

\begin{figure*}
\vspace{1.2cm}
  \includegraphics[width=\textwidth]{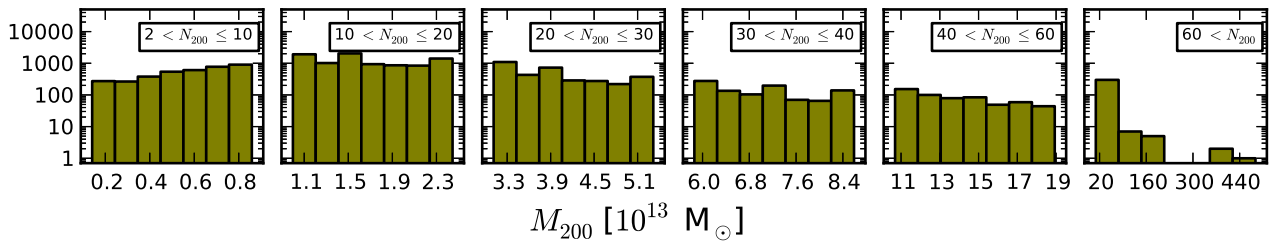}
  \caption{The underlying distribution of cluster candidate masses, within each of the six richness bins in Figure \ref{plot:nbinned}, for the full miscentred model. Because the parameters fit to the shear measurements are the normalization of the mass-richness relation (Equation \ref{MNeq}) and the miscentring parameters $p_{\mathrm{cc}}$ and $\sigma_{\mathrm{off}}$, the full ({\it not binned}) set of cluster $N_{200}$ values are each converted to an individual cluster mass. We bin these masses for presentation in the above histograms only, but emphasize that the composite-halo modelling approach in this work treats every cluster candidate as having an individual mass (richness) and redshift. This figure is also a visual representation of the 3D-MF cluster mass function, as obtained from weak lensing shear.}
\label{plot:multimass}
\end{figure*}

\begin{table*}
    \caption{Details of the ``Full Model'' fits for the richness-binned measurements (Equation \ref{modelEQ}, green curves in Figure \ref{plot:nbinned}). This model has 7 degrees of freedom. We list the richness range selected, the number of cluster candidates in that bin, the shear detection significance, and the average richness and redshift of clusters in the bin. Fitted parameters include the centring-related parameters $p_{\mathrm{cc}}$ and $\sigma_{\mathrm{off}}$, and the normalization of the mass-richness relation $M_0$, from which the average mass in each bin $\langle M_{200} \rangle$ is derived. Note that the average mass given is not the value fit itself, but the average of all resulting masses fit using the composite-halo approach discussed in Section \ref{sec:nfw}. See Figure \ref{plot:multimass} for a summary of the mass distributions within each $N_{200}$ bin. Reduced generalized $\chi^2$ are given for each bin, and should be compared with the corresponding fits listed in Table \ref{richbintable2}, for the simple one-parameter model assuming perfect centres.}
    \begin{tabular}{llllllllll}
      \hline
      Richness & Clusters & Significance & $\langle N_{200} \rangle$ & $\langle z_l \rangle$ & $p_{\mathrm{cc}}$ & $\sigma_{\mathrm{off}}$ & $M_0 \left[ 10^{13} M_{\odot}\right]$ & $\langle M_{200} \rangle \left[ 10^{13} M_{\odot}\right]$ & $\chi^2_{\mathrm{red}}$ \\ \hline
      2 $<N_{200}\leq$ 10 & 3745 & 14.2$\sigma$ & 8 & 0.45 & 1.0$_{-0.2}$ & \---- & $2.4^{+0.9}_{-1.0}$ & $0.6^{+0.2}_{-0.3}$ & 2.1  \\
      10 $<N_{200}\leq$ 20 & 9034 & 22.8$\sigma$ & 15 & 0.63 & 0.5$\pm$0.1 & $(0.40^{+0.06}_{-0.2})'$ & $2.4\pm0.6$ & 1.6$\pm$0.4 & 2.3 \\
      20 $<N_{200}\leq$ 30 & 3409 & 25.6$\sigma$ & 24 & 0.67 & 0.5$\pm$0.1 & $(0.4^{+0.2}_{-0.1})'$ & $2.9\pm0.5$ & 3.9$\pm$0.7 & 0.8 \\
      30 $<N_{200}\leq$ 40 & 986 & 23.4$\sigma$ & 35 & 0.65 & 0.5$\pm$0.2 & $(0.4\pm0.1)'$ & $3.0\pm0.7$ & 7$\pm$2 & 2.6 \\
      40 $<N_{200}\leq$ 60 & 568 & 22.2$\sigma$ & 48 & 0.60 & 0.54$\pm$0.08 & $(1.3^{+0.5}_{-0.4})'$ & $3.6^{+0.8}_{-1.0}$ & $14^{+3}_{-4}$ & 0.3 \\
      60 $<N_{200}$ & 314 & 22.5$\sigma$ & 114 & 0.55 & 0.5$\pm$0.2 & $(0.4^{+0.2}_{-0.1})'$ & $1.6^{+0.4}_{-0.5}$ & 26$^{+6}_{-7}$ & 3.4 \\
      \hline
    \end{tabular}
\label{richbintable1}
\end{table*}

\begin{table*}
    \caption{This table is a companion to Table \ref{richbintable1}, giving details of the $p_{\mathrm{cc}} \equiv 1$ model fits for the richness-binned measurements (purple dashed curves in Figure \ref{plot:nbinned}). This model has 9 degrees of freedom. We list the richness range selected (the reader can refer to Table \ref{richbintable1} for the number of clusters, shear significance, and average richness and redshift). For this model, there is a single fit parameter, the normalization of the mass-richness relation $M_0$, from which $\langle M_{200} \rangle$ is derived (again see Figure \ref{plot:multimass} for the full distribution of masses in each richness bin).}
    \begin{tabular}{llll}
      \hline
      Richness & $M_0 \left[ 10^{13} M_{\odot}\right]$ & $\langle M_{200} \rangle \left[ 10^{13} M_{\odot}\right]$ & $\chi^2_{\mathrm{red}}$ \\ \hline
      2 $<N_{200}\leq$ 10 & 2.4$^{+0.4}_{-0.6}$ & 0.6$\pm$0.1 & 1.6  \\
      10 $<N_{200}\leq$ 20 & 1.8$\pm$0.2 & 1.2$\pm$0.2 & 4.8  \\
      20 $<N_{200}\leq$ 30 & 2.2$^{+0.2}_{-0.3}$ & 3.0$^{+0.3}_{-0.4}$ & 5.3  \\
      30 $<N_{200}\leq$ 40 & 2.4$\pm$0.3 & 5.5$\pm$0.8 & 4.4  \\
      40 $<N_{200}\leq$ 60 & 2.1$\pm$0.3 & 8$\pm$1 & 4.7  \\
      60 $<N_{200}$ & 1.4$\pm$0.2 & 23$\pm$3 & 4.4  \\
      \hline
    \end{tabular}
\label{richbintable2}
\end{table*}

\subsection{Fits to $\Delta\Sigma$}
\label{fits}

\begin{figure*}
\vspace{0.5cm}
  \includegraphics[width=0.6\textwidth]{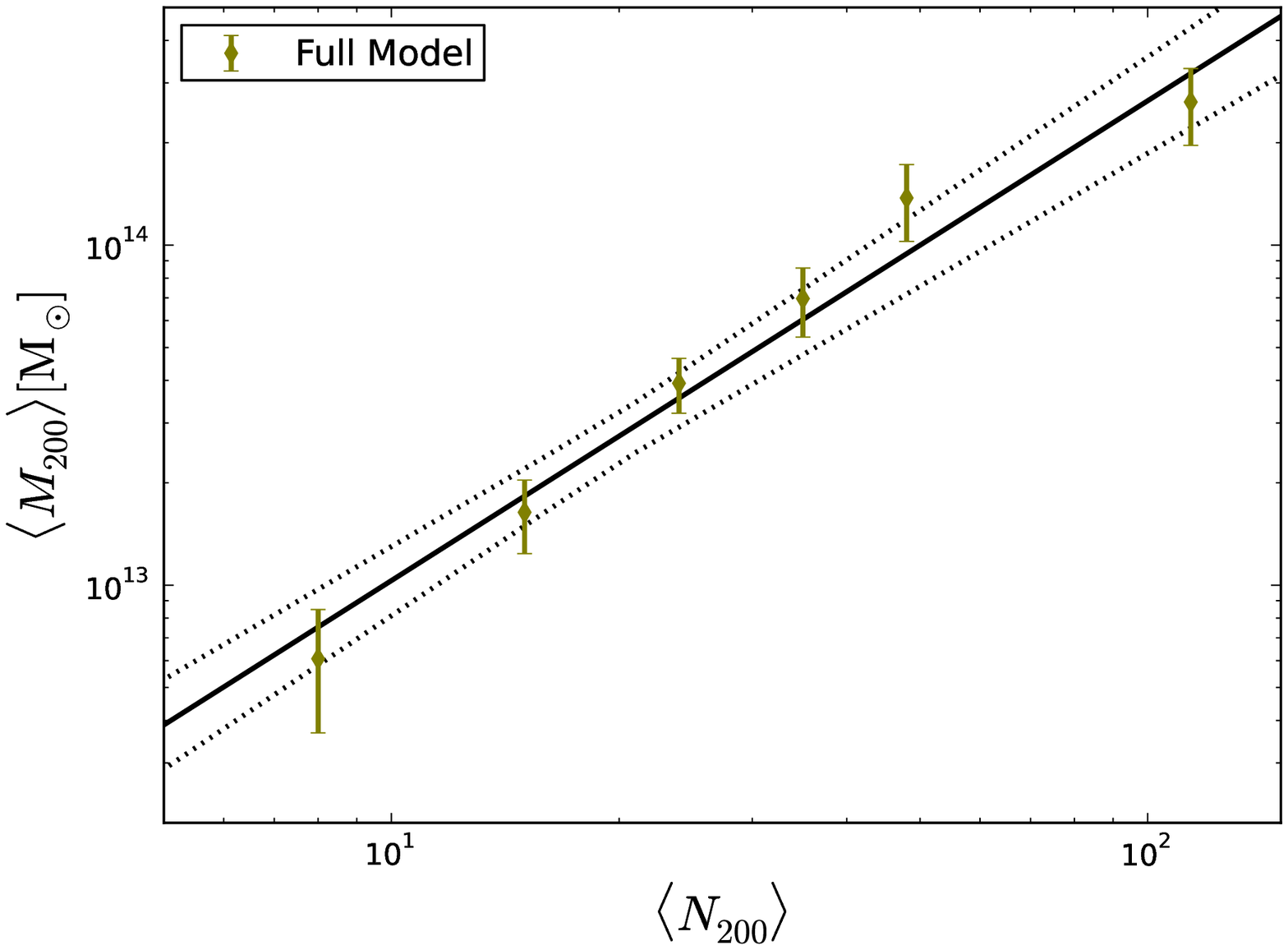}
  \caption{Power law best fit to mass-richness relation (Equation \ref{MNeq}) obtained from average masses measured for the individual $N_{200}$ bins in Figure \ref{plot:nbinned} and Table \ref{richbintable1}, for the full model which accounts for miscentring, and including the (very small) correction for intrinsic scatter. The dotted lines show the 1$\sigma$ limits on this relation. As discussed in Section \ref{sec:MN} the simple $p_{\mathrm{cc}} \equiv 1$ model, which assumes perfect cluster centres, yields the same slope, but a slightly lower overall normalization.}
\label{plot:massrich}
\end{figure*}

We divide our cluster candidate catalogue into six richness bins, and measure the differential surface mass density as described in Section \ref{sec:measure}. The significances of the separate stacked measurements of $\Delta\Sigma(R)$ shown in Figure \ref{plot:nbinned} range from 14.2$\sigma$ to 25.6$\sigma$, calculated using the full covariance matrices to include correlation between radial measurement bins. Error bars are calculated as the square root of the diagonal of the covariance matrices. These values, along with details of the richness bins and fits are given in Table \ref{richbintable1}. This yields a total 3D-MF cluster shear significance of $\sim$54$\sigma$. 

In modelling the halo mass, we use a composite-halo approach, which allows for the fact that the cluster candidates in a given stacked measurement may have a range of individual masses and redshifts. We emphasize that instead of fitting a single average mass (and also avoiding a single effective cluster redshift), we actually fit to the normalization of the mass-richness relation, $M_0$. We convert the array of cluster $N_{200}$ values into masses with the equation
\begin{equation}
\label{MNeq1.5}
M_{200} = M_0 \left( \frac{N_{200}}{20} \right)^{1.5}.
\end{equation}

In each separate stacked weak lensing measurement, we keep the slope of this mass-richness relation fixed, to avoid over-fitting to each stack with parameters that are quite degenerate within a narrow cluster bin. The NFW mass of each individual cluster is given by Equation \ref{MNeq1.5}, with the fixed slope of 1.5 from \citet{Ford14}, which will be shown to be consistent with the global mass-richness relation, measured and discussed in Section \ref{sec:MN} of this current work. We note that because of the free normalization $M_0$, this approach does neither impose the form of the richness distribution (Figure \ref{plot:hists}) nor does it set a prior on the individual mass.

We fit the halo model given in Equation \ref{modelEQ} to the data, employing the downhill simplex method to minimize the generalized $\chi^2$, using the full covariance matrices estimated from bootstrap resampling. The results are displayed as the green curves in Figure \ref{plot:nbinned} (labelled ``Full Model''), and summarized in Table \ref{richbintable1}. The number of degrees of freedom for the model is 7 (10 radial bins minus 3 fit parameters). 

To emphasize the importance of cluster miscentring, we also plot the best-fitting model where $p_{\mathrm{cc}} \equiv 1$ (i.e. perfect cluster centres) for comparison. This is shown as the dashed purple curves in Figure \ref{plot:nbinned} (with a single fit parameter, $M_0$, this model has 9 degrees of freedom). Visual inspection reveals poor fits to the data at small radii for this model, and this fact is quantified by the reduced generalized $\chi^2$ statistic ($\chi^2_{\mathrm{red}}$) values in Table \ref{richbintable2}. These results imply that cluster centroiding is an important component in the modelling of the 3D-MF weak lensing shear mass profiles, especially at the high mass (richness) end. For the majority of the rest of this work we will focus our attention on the results of the full model, which accounts for offset cluster centres.

The ensemble of cluster masses that result from the composite-halo modelling approach are displayed in Figure \ref{plot:multimass}, where each panel represents a single stacked weak lensing measurement, congruent with Figure \ref{plot:nbinned}. This visual representation of the cluster mass function is largely distinct from the $N_{200}$ histogram in Figure \ref{plot:hists}, because these masses are dependent upon the mass-richness normalization, as well as the miscentring parameters, which are fit to the measurements.

\begin{figure*}
\vspace{1cm}
  \includegraphics[width=0.9\textwidth]{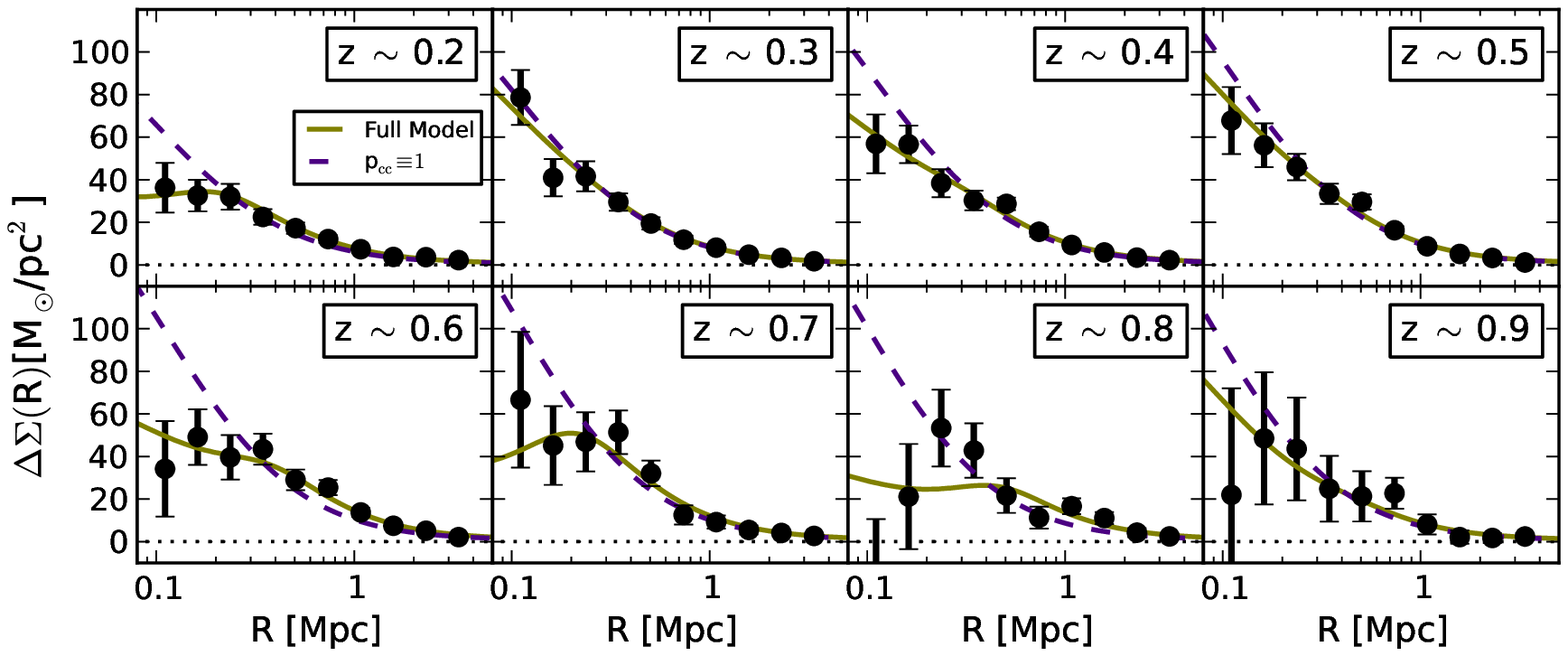}
  \caption{Best-fitting models for each stack of cluster candidates, this time binned in redshift. As in Figure \ref{plot:nbinned}, the solid green curves are the best fits to the full model given by Equation \ref{modelEQ}. The dashed purple curves are the best-fitting models which assumes that every cluster centre identified by 3D-MF is perfectly aligned with the dark matter halo centre. Tables \ref{ztable1} and \ref{ztable2} summarize the results of these fits.}
\label{plot:zbinned}
\end{figure*}

\begin{table*}
    \caption{Details of the ``Full Model'' fits for the redshift-binned measurements (green curves in Figure \ref{plot:zbinned}). This model has 7 degrees of freedom. We list the same bin properties and fits given in Table \ref{richbintable1}. The systematic errors listed on some cluster masses stem from uncertainties on the exact redshift of the cluster candidate. The fits in this table should be compared with the corresponding values in Table \ref{ztable2}, which represents the perfectly centred model.}
    \begin{tabular}{lllllllll}
      \hline
      Redshift & Clusters & Significance & $\langle N_{200} \rangle$ & $p_{\mathrm{cc}}$ & $\sigma_{\mathrm{off}}$ & $M_0 \left[ 10^{13} M_{\odot}\right]$ & $\langle M_{200} \rangle \left[ 10^{13} M_{\odot}\right]$ & $\chi^2_{\mathrm{red}}$ \\ \hline
      $z$ $\sim$ 0.2 & 1161 & 13.8$\sigma$ & 14 & 0.3$\pm$0.3 & $(0.4^{+0.3}_{-0.1})'$ & 3$\pm$1 & 2.3$^{+0.9}_{-1.0}\pm$0.4$^{\mathrm{sys}}$ & 0.6 \\
      $z$ $\sim$ 0.3 & 1521 & 15.7$\sigma$ & 17 & 0.8$^{+0.2}_{-0.3}$ & $(0.4^{+1}_{-0.4})'$ & 2.3$^{+0.7}_{-0.9}$ & 2.6$^{+0.8}_{-0.9}\pm$0.2 & 0.4 \\
      $z$ $\sim$ 0.4 & 2248 & 17.0$\sigma$ & 18 & 0.7$\pm$0.2 & $(0.4^{+0.3}_{-0.2})'$ & 2.6$\pm$0.9 & 3$\pm$1$\pm$0.1$^{\mathrm{sys}}$ & 0.8 \\
      $z$ $\sim$ 0.5 & 2935 & 20.2$\sigma$ & 18 & 0.8$\pm$0.2 & $(0.4^{+0.2}_{-0.3})'$ & 2.5$^{+0.6}_{-0.8}$ & 3.0$^{+0.7}_{-1.0}$ & 1.7 \\
      $z$ $\sim$ 0.6 & 2456 & 14.7$\sigma$ & 20 & 0.4$\pm$0.2 & $(0.4\pm0.1)'$ & 3$\pm$1 & 4$\pm$1 & 1.1 \\
      $z$ $\sim$ 0.7 & 2331 & 11.9$\sigma$ & 22 & 0.7$\pm$0.3 & $(0.4^{+0.6}_{-0.4})'$ & 2.1$^{+0.9}_{-1.0}$ & 3$\pm$1 & 0.8 \\
      $z$ $\sim$ 0.8 & 2364 & 8.7$\sigma$ & 22 & 0.2$\pm$0.2 & $(0.4\pm0.2)'$ & 3$^{+1}_{-3}$ & 4$^{+2}_{-3}$ & 1.9 \\ 
      $z$ $\sim$ 0.9 & 3040 & 6.8$\sigma$ & 19 & 0.6$\pm$0.4 & $(0.4^{+1}_{-0.4})'$ & 1.8$^{+0.8}_{-1.7}$ & 1.9$^{+0.9}_{-1.8}$ & 0.5 \\
      \hline
    \end{tabular}
    \label{ztable1}
\end{table*}

\begin{table*}
    \caption{This table is a companion to Table \ref{ztable1}, giving details of the $p_{\mathrm{cc}} \equiv 1$ model fits for the redshift-binned measurements (purple dashed curves in Figure \ref{plot:zbinned}). This model has 9 degrees of freedom. For this model, there is a single fit parameter, the normalization of the mass-richness relation $M_0$, from which $\langle M_{200} \rangle$ is derived.}
    \begin{tabular}{llll}
      \hline
      Redshift & $M_0 \left[ 10^{13} M_{\odot}\right]$ & $\langle M_{200} \rangle \left[ 10^{13} M_{\odot}\right]$ & $\chi^2_{\mathrm{red}}$ \\ \hline
      $z$ $\sim$ 0.2 & 2.6$\pm$0.6 & 2.0$\pm$0.5$\pm$0.3$^{\mathrm{sys}}$ & 2.1  \\
      $z$ $\sim$ 0.3 & 2.1$\pm$0.4 & 2.4$\pm$0.4$\pm$0.2$^{\mathrm{sys}}$ & 0.4  \\
      $z$ $\sim$ 0.4 & 2.2$\pm$0.4 & 2.7$\pm$0.5$\pm$0.1$^{\mathrm{sys}}$ & 1.4  \\
      $z$ $\sim$ 0.5 & 2.2$\pm$0.3 & 2.7$\pm$0.4 & 1.6  \\
      $z$ $\sim$ 0.6 & 2.4$\pm$0.6 & 2.9$\pm$0.7 & 4.5  \\
      $z$ $\sim$ 0.7 & 1.9$^{+0.4}_{-0.5}$ & 2.4$^{+0.6}_{-0.7}$ & 0.8  \\
      $z$ $\sim$ 0.8 & 1.4$\pm$0.6 & 1.8$\pm$0.8 & 3.3  \\ 
      $z$ $\sim$ 0.9 & 1.3$\pm$0.6 & 1.4$\pm$0.6 & 0.6  \\
      \hline
    \end{tabular}
    \label{ztable2}
\end{table*}


\subsection{The Mass-Richness Relation}

\label{sec:MN}
The results of the previous section demonstrate a strong scaling of mass with richness. In Figure \ref{plot:massrich} we plot the average mass $M_{200}$ measured in each richness bin, as a function of richness $N_{200}$, and fit the power law scaling relation:
\begin{equation}
\label{MNeq}
M_{200} = M_0 \left( \frac{N_{200}}{20} \right)^\beta.
\end{equation}
This is similar to Equation \ref{MNeq1.5}, but the slope $\beta$ is now a free parameter, and the mass-richness normalization $M_0$ is fit across the full distribution of clusters. We note that the choice of $\beta=1.5$ in Equation \ref{MNeq1.5} does not have a significant effect on the $\beta$ measured here. Because of the degeneracy between $\beta$ and $M_0$ in each narrow cluster bin, a different choice of slope for the measurements in Section \ref{fits} still yields essentially the same mass estimates $M_{200}$, and thus the same global mass-richness relation.

Since galaxy clusters exhibit a natural intrinsic scatter between halo mass and richness (or other mass proxy), a bias in scaling relations can result if this scatter is ignored \citep{Rozo09a}. The idea here is that while galaxy clusters at a given richness will scatter randomly with regard to their average mass, because of the shape of the cluster mass function, the net effect is to scatter from low to high mass. This can lead to a biased mass estimate in a given richness bin, as well as affect the global result for the mass-richness relation. We correct for intrinsic scatter using the data itself, following a procedure inspired by \citet{Velander14}, which is as follows.

We first fit Equation \ref{MNeq} to the uncorrected raw mass estimates from each richness bin, and use this power law relation to assign an individual mass to each cluster, based on its value of $N_{200}$. We then draw many ``simulated'' clusters from the observed cluster mass function (i.e. the $N_{200}$ histogram in Figure \ref{plot:hists}), taking 1000 times as many ``simulated'' as observed clusters. We then scatter their masses by values drawn from a Gaussian in $\ln (M_{200})$, with width $\sigma_{\ln M|N}$, centred on the particular $N_{200}$. For the width of the intrinsic scatter, we use values estimated by \citet{Rozo09a} for the MaxBCG clusters in the Sloan Digital Sky Survey (SDSS). This is $\sigma_{\ln M|N} \sim 0.45$, which is the scatter in the natural logarithm of mass, at fixed richness.

The resulting mass estimates are then used to calculate the corrected arithmetic mean mass in each of the richness bins, which are plotted in Figure \ref{plot:massrich} and used to re-fit Equation \ref{MNeq}, yielding the final mass-richness relation reported below. The corrections applied to the mass estimates are at the sub-percent level, and therefore negligible compared to other sources of uncertainty in this work. Nevertheless, we include these small corrections when fitting for the mass-richness relation. We note that increasing $\sigma_{\ln M|N}$ up to the 95\% confidence limit reported by \citet{Rozo09a} still does not affect the conclusions drawn in this work. A glance at Figure \ref{plot:multimass} justifies the low-impact of the intrinsic scatter correction, as most richness bins do not exhibit a very strong slope, which would otherwise lead to a larger effect on average mass in each bin.

In this work we measure $M_0$ = $(2.7^{+0.5}_{-0.4}) \times 10^{13} M_{\odot}$ and $\beta$ = $1.4 \pm 0.1$ for the full model (Figure \ref{plot:massrich}), with a $\chi^2_{\mathrm{red}}$ of 0.9. For the perfectly centred model, we get $M_0$ = $(2.2 \pm 0.2) \times 10^{13} M_{\odot}$ and $\beta$ = $1.4 \pm 0.1$, with a $\chi^2_{\mathrm{red}}$ of 1.0. (Note that uncertainties are larger on parameters estimated from the full model, both here and throughout this work, since there are simply more parameters than the perfectly-centred model). These results demonstrate that not including the centroid uncertainty in our analysis would lead us to systematically underestimate the cluster masses as well as the mass-richness normalization. Section \ref{sec:magn} contains a thorough comparison of these results with our previous magnification measurements of these cluster candidates.


\begin{figure}
\vspace{0.2cm}
  \includegraphics[scale=0.4]{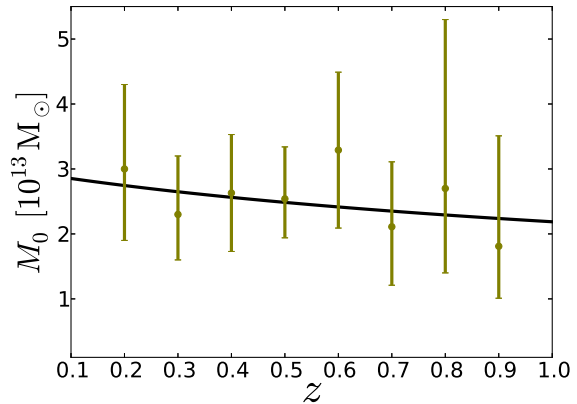}
  \caption{Normalization of the mass-richness relation $M_0$ as a function of redshift $z$. The evidence for redshift evolution is not significant: the mildly negative slope is consistent with zero.}
\label{plot:MoVSz} 
\end{figure}

\subsection{Results of Binning Clusters in Redshift}
We also investigate the weak lensing shear measurement of 3D-MF cluster candidates as a function of cluster redshift. 3D-MF sorts candidate clusters into bins of width $\Delta z \sim$ 0.1, so these are natural bin choices, and the same used in our previous analysis \citep{Ford14}. Figure \ref{plot:zbinned} shows the measurements and fits to $\Delta \Sigma$, with error bars again obtained from the covariance matrices (Section \ref{sec:measure}). The significance of the shear measurements reaches $\sim$ 20$\sigma$ at $z \sim$ 0.5, where there is an abundance of 3D-MF cluster candidates, and drops to $\sim$ 7$\sigma$ at the highest redshifts, where shear signal-to-noise is depleted. 

In Figure \ref{plot:zbinned} (similar to Figure \ref{plot:nbinned}), we plot the full model in solid green, and the perfectly centred model in dashed purple. Table \ref{ztable1} and Table \ref{ztable2} display the results and fit parameters for these two models, respectively. The measurements at lower redshifts have an additional systematic error listed, which stems from uncertainties on the cluster redshifts, due to the way the 3D-MF method slices in redshift space \citep{Ford14}. The 3D-MF cluster candidates are found to be quite similar in average mass across the range of redshift probed -- we consistently obtain measurements of a few $10^{13} M_{\odot}$. The best-fitting miscentring parameter $p_{\mathrm{cc}}$ varies somewhat erratically as a function of redshift, but the error bars are too large to infer any significance from this. The width of the offset distribution on the other hand remains squarely at $\sigma_{\mathrm{off}} \sim$ 0.4 arcmin. We discuss this result in relation to other cluster miscentring studies in Section \ref{sec:centres}.

We investigate possible redshift evolution of the mass-richness relation (given by Equation \ref{MNeq}) in Figure \ref{plot:MoVSz}, which shows the normalization of this scaling relation, $M_0$, as a function of redshift (with $\beta$ = 1.5 fixed), as listed in Table \ref{ztable1}. We fit a powerlaw relation of the form
\begin{equation}
\label{MoZeq}
M_0 (z) = M_0 (z=0) \cdot \left[ 1+z \right] ^{\gamma}.
\end{equation}
We find a normalization $M_0 (z=0) = (3.0 \pm 0.6) \times 10^{13} M_{\odot}$, and a powerlaw slope $\gamma = -0.4^{+0.5}_{-0.6}$. The slope is consistent with zero, so no significant redshift-evolution is detected for the 3D-MF mass-richness scaling relation.


\section{Discussion}


\subsection{Interpretation of the Results}

The 3D-MF clusters represent a wide range of halo masses and impose a significant shear signal on background galaxies. The measured $\Delta \Sigma$ profiles from different stacked subsamples of clusters yield an important glimpse at the state of the dark matter haloes. We fit a model that includes parameters designed to distinguish the fraction of well-centred versus offset haloes, and the width of the offset distribution. The latter is consistently measured to peak at an offset of $\sim$ 0.4 arcmin, except for the richness bin 40 $< N_{200} \leq$ 60, for which we find a larger best fit of 1.3 arcmin (this much larger offset is puzzling, and will require follow-up to determine whether it is physical or perhaps a spurious effect of overfitting). The fraction of clusters that are not correctly centred is generally about 50\% across richness bins, but has large error bars that do not allow us to distinguish interesting features at a statistically significant level. Nonetheless, we do find overall that the 3D-MF cluster halo profiles are better fitted by not enforcing perfect centroiding.

This study comprises several novel components, which will be discussed in more detail below. The large number of clusters, and the fact that 3D-MF does not assume anything about cluster galaxy colours, makes the uniqueness of the data set valuable in its own right.  Evolution of the normalization of the mass-richness relation across a wide span of redshift has only been constrained previously by \citet{EdoThesis12} and \citet{Andreon14}. The direct comparison between shear and magnification measured masses is a first for a cluster catalogue of this volume. There are several caveats to the implications of this work, notably the very likely presence of false-detections at the low-significance (low-richness) end of the cluster candidate spectrum.


\subsection{Comparisons of Cluster Catalogue Volume}

The most noteworthy aspect of the CFHTLS-Wide 3D-MF cluster catalogue is its sheer size. With over 100 cluster candidates per square degree (18056 clusters in 154 deg$^2$), spanning redshifts up to $z \sim$ 0.9, this compilation of cluster candidates is one of the most complete available. We encourage others to utilize this catalogue, available from \url{cfhtlens.org}, as there are an abundance of scientific investigations now possible with it. 

The current widest survey with a galaxy cluster catalogue is SDSS. The SDSS collaboration found 13823 galaxy groups and clusters spread over 7500 deg$^2$, using their maxBCG method \citep{Koester07}. This amounts to less than two clusters per deg$^2$, and is restricted to lower redshifts (0.1 $< z <$ 0.3). The maxBCG technique relies on finding potential bright galaxies and searching around them for the presence of a red sequence in colour-magnitude space (which would indicate the presence of red, elliptical galaxies, common in galaxy clusters). 

Interestingly, visual inspection of 3D-MF galaxy cluster candidates shows that the lower redshift clusters often do have a bright central galaxy, but this is less true at higher redshifts. It would be interesting to quantify this aspect in future work, especially when an opportunity presents itself to compare 3D-MF to other algorithms directly, by running both on the same optical data set. Galaxy clusters do not always have one brightest central galaxy, and if they do have one, it is not always exactly in the centre of the galaxy cluster, so comparing the biases of both methods could ultimately result in a more complete cluster list, or could potentially show the limitations of methods like maxBCG. 

Several cluster catalogues have been compiled in the CFHTLS-Wide. \citet{Durret11} used photometric redshift information to construct galaxy density maps in CFHTLS, building upon earlier work by \citet{Adami10} and \citet{Mazure07}. They found 4061 cluster candidates in the Wide fields, with masses greater than about 10$^{14} M_{\odot}$, spanning redshifts 0.1 $\le z \le$ 1.15. \citet{Shan12} used a 3D-lensing approach, with convergence maps and galaxy photometric redshifts, to detect 85 clusters at $\langle z \rangle \sim$ 0.36 in the W1 field of the CFHTLS-Wide.

\citet{Wen12} compiled an optical cluster catalogue from SDSS-III, using galaxy photometric redshifts and a friends-of-friends algorithm. They found an impressive 132684 clusters over 14000 deg$^2$ in a redshift range 0.05 $\lesssim z <$ 0.8. A recent cluster shear analysis was done by \citet{Covone14}, using the overlapping portion of the \citet{Wen12} catalogue, with the CFHTLenS shear catalogue. To date, this is the most complete cluster catalogue analysed in the context of CFHTLenS, but still the cluster density is $\lesssim$ 1/10th of that achieved with 3D-MF. A comparison of 3D-MF with the cluster catalogues compiled using these different techniques will be presented in a future analysis.


\subsection{Comparison with other Mass-Richness Relations}

The 3D-MF cluster finder presents us with a sample of cluster candidates which, like every other cluster-finder, are drawn from a somewhat unique distribution defined by its particular selection function. Despite the difficulties inherent to making exact comparisons between scaling relations measured on disparate cluster samples, we attempt a broad look at how the 3D-MF mass-richness scaling compares to other relations in the literature.

\citet{Wen09} defined a measure of richness $R$ for their SDSS clusters, which is somewhat similar to the $N_{200}$ used in this work. They counted all galaxies brighter than absolute magnitude $M_r \leq -21$, within a 1 Mpc radius and $\Delta z < 0.04(1+z)$. Converting their mass-richness relation to the form of ours (Equation \ref{MNeq}), they obtained a somewhat steeper slope $\beta \sim$ 1.9, and a higher normalization $M_0 \sim 2.5\times10^{14} M_{\odot}$ than the best-fitting models presented in this work ({\it Full Model}: $M_0 \sim 2.7\times10^{13} M_{\odot}$, $\beta \sim$ 1.4). We tried measuring richness for the 3D-MF clusters following the same prescription as \citet{Wen09}, but found the one-size-fits-all radius to be a serious limitation for our sample, since the 3D-MF cluster candidates span a wide range of masses and therefore characteristic radii. The resulting richness estimates had greatly enhanced scatter and did not scale well with mass at the more massive end of the cluster catalogue.

In a follow-up paper, \citet{Wen12} defined a new richness $R_{L*}$ -- the total r-band luminosity within $R_{200}$ in units of $L_{\odot}$. For the portion of clusters with previously measured masses (weak lensing or X-ray), a scaling between the radius $R_{200}$ derived from these masses and the luminosity within 1 Mpc was measured, and this was used to estimate radii for calculating $R_{L*}$ for the full sample of 132684 clusters. For the subsample with existing mass estimates, \citet{Wen12} found a mass-richness relation with normalization $M_0 \sim 1.1 \times 10^{14} M_{\odot}$ and slope $\beta \sim$ 1.2 (again converting to the form of our Equation \ref{MNeq}). \citet{Covone14} measured weak lensing masses for 1176 of the clusters from \citet{Wen12}, which overlapped with CFHTLenS. They found a very similar mass-richness scaling, with $M_0 \sim 10^{14} M_{\odot}$ and $\beta \sim$ 1.2. 

The mass-richness slope of the 3D-MF cluster candidates sits squarely between the results of \citet{Wen09}, using the $R$ richness, and \citet{Wen12} and \citet{Covone14}, which used the $R_{L*}$ measure. The 3D-MF normalization is lower than the other cluster catalogues, which could partly be a result of 3D-MF detecting more lower mass clusters missed by other finders. However, the different definition of richness, namely the fainter limit on galaxies contributing to $N_{200}$, means that the same mass cluster will have a larger measured richness in this work, implying a lower mass-richness normalization. Finally, the presence of false detections in the 3D-MF catalogue (estimated from simulations to be at the level of $16-24$\%) would certainly bias the mass estimates low.

\citet{Johnston07} used a quite different definition of richness for the maxBCG clusters, counting only red-sequence galaxies brighter than 0.4$L_*$, within an $R_{200}^{\rm gals}$ that was estimated from the number of galaxies within 1 Mpc \citep[following a prescription in][]{Hansen05}. Weak lensing masses were used to find a normalization $M_0 \sim 1.3 \times 10^{14} M_{\odot}$ and slope $\beta \sim$ 1.3. \citet{Rozo09b} created updated richness estimates of the maxBCG clusters by applying an improved colour modelling of cluster members, and allowing individual cluster radii to vary until the scatter between richness and X-ray luminosity was minimized.

\citet{Andreon10} defined a measure of richness for the Cluster Infall Regions in SDSS catalogue, for which masses $M_{200}$ and radii $R_{200}$ were already available (from application of the caustic technique). They studied a sample of 53 low-redshift clusters, in the range $0.03 < z < 0.1$, and their $N_{200}$ included all red galaxies brighter than $M_V = -20$ within the radius $R_{200}$. In a follow-up analysis they measured a tight mass-richness scaling relation with normalization $M_0 \sim 1.4 \times 10^{11} M_{\odot}$ and slope $\beta \sim 2.1$ \citep{Andreon12}.

The addition of the galaxy colour information in the richness estimate of the previous three examples, in particular, creates difficulty in drawing meaningful comparisons between their mass-richness scaling relation and the 3D-MF scaling relation. We emphasize that the value of any mass-richness relation is limited to the particular cluster sample for which it was derived, which in turn depends on the cluster-finding algorithm and details of the survey on which the catalogue was compiled. As discussed in \citet{Rozo09b}, the simple fact that estimates of richness are readily available in an optical cluster survey, and that they can be applied to clusters of virtually any mass, nevertheless makes richness a worthwhile parameter to measure. So although richness has many different definitions, and some unavoidable scatter in its scaling relations with various cluster mass estimates, it remains a useful tool for characterizing galaxy clusters. 


\subsection{Comparisons with other Cluster Centroid Analyses}
\label{sec:centres}
We find the distribution of centroid offsets to be well characterized by a Gaussian of width $\sigma_{\mathrm{off}} \sim$ 0.4 arcmin,\footnote[2]{For comparisons, 0.4 arcmin $\sim$ 147 kpc at redshift 0.5} and that this miscentring has an effect on a significant portion of the candidate clusters (up to $\sim$ 80\% of them are affected, see Table \ref{richbintable1}). Interestingly, previous studies applying 3D-MF to simulations yielded an average $\sigma_{\mathrm{off}}=0.40 \pm 0.06$ arcmin \citep[see Figure 1 in][]{Ford14}, which is easily consistent with the best-fitting offset measured on the real 3D-MF cluster candidates in this work.

\begin{figure*}
\vspace{1cm}
  \includegraphics[width=0.9\textwidth]{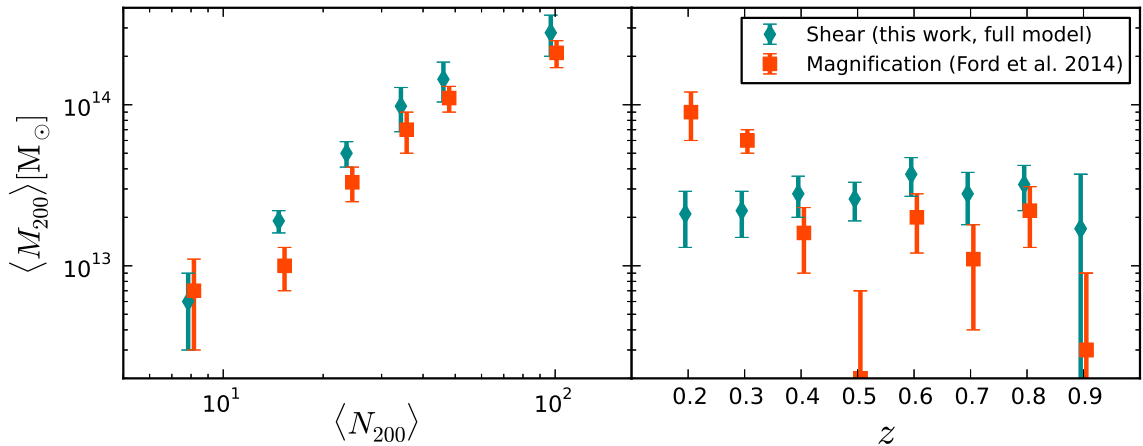}
  \caption{Here we compare the mass measurements obtained for the 3D-MF cluster candidates using weak lensing shear (i.e. this work) with the results obtained measuring the masses with the lensing magnification technique \citep[the $N_{200}$ estimates from that work, ][are used in this plot for the purposes of comparison]{Ford14}. The first panel compares mass measurements when cluster candidates are binned in richness $N_{200}$, and the second panel shows the redshift $z$ binning. Bins are identical for magnification and for shear, but the points are slightly offset horizontally for clarity. Blue diamonds represent the shear, and orange squares are for magnification.}
\label{plot:magshear}
\end{figure*}

The maxBCG clusters were found to have centroid offsets around 0.42 $h^{-1}$Mpc, based on simulations \citep{Johnston07}, which is several times larger than the ones measured for the 3D-MF cluster candidates. There were large uncertainties associated with the probability of a cluster having a correct centroid selected, but this was determined to be approximately $\ge$50\% \citep[see Figure 5 in][]{Johnston07}, which is similar to $p_{cc}$ found in this work. \citet{George12} performed a miscentring analysis of X-ray groups in the COSMOS field. They found offsets of $\sim 20-70$ kpc, for different candidate centres, which are smaller than measured for 3D-MF clusters. 

It is worth noting that candidate cluster centres that are coincident with a member galaxy have been found to better trace the halo's centre of mass, relative to other types of centroids such as X-ray, or various weighted centres of galaxy positions \citep{George12}. See also \citet{Bildfell08} for a study of massive X-ray clusters. 3D-MF centres (peaks in the likelihood map) do not necessarily coincide with a cluster galaxy member, so future work should investigate various possible candidate centres to find the one that best traces the centre of mass for 3D-MF cluster haloes.


\subsection{Comparison with Magnification Results}
\label{sec:magn}

One of the most interesting aspects of this work is the direct comparison between magnification and shear mass estimates, which is now made possible in the context of a very large lens sample. Prior to this work, the only observational magnification--shear direct cluster mass comparison in the literature was \citet{Ford12}. That study demonstrated a 1$\sigma$ consistency between masses measured with the two techniques, but applied to a small sample of just 44 galaxy groups, so any trends in cluster size or redshift were unable to be explored. \citet{Huff14} compared magnification and shear masses for SDSS galaxy lenses, using a different and novel approach to measuring lensing magnification, and found mass profiles to be within a factor 3 of agreement.

Important work related to the {\it joint} analysis of shear and magnification has been developed in \citet{Umetsu11} and \citet{Umetsu13}. \citet{Umetsu14} combined shear and magnification to measure the mass profiles of 20 massive X-ray-selected clusters. This work demonstrated that the geometric mean mass of the shear+magnification measurement was consistent with the shear-only measurement, but did not show magnification results on their own. Earlier work in \citet{Umetsu11} compared the signal-to-noise of the magnification and shear, but did not present mass estimates from separate analyses. 

In this work we exploit the volume of the 3D-MF cluster catalogue to fully compare masses determined with each of the independent techniques, as a function of both candidate cluster richness and redshift. The findings are summarized in Figure \ref{plot:magshear}. For consistency in the comparison, this plot uses the original $N_{200}$ estimates from \citet{Ford14}, so that the cluster candidate stacks in each richness bin are identical. Also, in this section only, we use the mass-concentration relation of \citet{Prada12}, in identical fashion to the magnification work. The \citet{Prada12} relation is in excellent agreement with recent measurements by \citet{Covone14} of the masses and concentrations of a cluster sample in CFHTLenS, although it is in tension with other measurements such as \citet{Merten14}. 

The left-hand panel of Figure \ref{plot:magshear} displays the results when 3D-MF cluster candidates are stacked across all redshifts. The average of the composite-halo masses fit to each stack is comparable between the two methods, but the magnification estimates are systematically lower than the shear estimates. This yields a mass-richness normalization which is about 2$\sigma$ higher for the shear method, although the slope of the relation recovered with the two techniques is essentially identical. The magnification measurements yielded $M_0 = (2.2 \pm 0.2) \times 10^{13} M_{\odot}$ and $\beta = 1.5 \pm 0.1$ \citep[see the miscentred model in][]{Ford14}, while the shear measurements here give $M_0 = (3.1 \pm 0.5) \times 10^{13} M_{\odot}$ and $\beta = 1.5 \pm 0.2$. We note that the mass-richness relation parameters obtained from the shear measurements in Figure \ref{plot:magshear} are consistent within 1$\sigma$ with the new mass-richness parameters obtained with shear in Figure \ref{plot:massrich} and discussed in Section \ref{sec:MN}. We reiterate that the slight difference between the shear measurements in Figures \ref{plot:massrich} and \ref{plot:magshear} is due to a recalibration of the cluster $N_{200}$ estimates (see Section \ref{3DMF}) and a different choice of mass-concentration relation.

It is important to note that in both of the aforementioned magnification studies \citep{Ford12,Ford14}, the background source sample is completely distinct from the background sources used to measure shear. Indeed, both magnification results used magnified Lyman-break galaxies, which are point-like sources whose negligible apparent size would not permit a measurement of the shear. In this sense, the magnification results are largely independent from the shear measurements to which they are compared, having only the lens population in common. We note that alternative methods of measuring magnification using source size information would instead tend to use the same source sample employed for measuring shear.

The comparison becomes more interesting as a function of redshift, shown in the right-hand panel of Figure \ref{plot:magshear}. Here we see that the shear-measured average mass of cluster candidates does not vary as a function of redshift, while the magnification masses fluctuate. In \citet{Ford14}, we discussed this behaviour of the magnification signal, but without an alternative mass determination were not able to conclude whether this variation communicated an intrinsic property of the 3D-MF cluster candidates, or was an artefact of the magnification measurement. We are still unable to say with certainty whether the masses of the 3DMF cluster sample truly are constant or evolving across the redshift range, as suggested by the conflicting shear and magnification measurements. Here, we discuss several possible reasons for these discrepant redshift-binned results. 

First of all, the distributions of richness values for the separate $z$ slices are very similar, with the lower redshift slices containing relatively higher fractions of low-richness cluster candidates \citep[see Figure 7 in][]{Ford14}. So if (1) richness is a good estimator for mass, which it appears to be given the strong scaling, and (2) the mass-richness relation does not evolve strongly with redshift over the range $z \sim 0.5 \rightarrow 0.2$, then we would expect similar masses across this range, or for masses to actually decrease at lower redshift in concordance with the lower mean cluster $N_{200}$ (i.e. the opposite of the trend suggested by magnification).

As discussed in detail in \citet{Ford14}, at $z \sim 0.2-0.3$ the magnification measurement is expected to be affected by some low-$z$ contamination in the Lyman-break galaxy source sample. In that work we attempted to compensate for this effect by including a term in the modelling of the measured signal, to account for physical clustering where the populations overlapped. A crucial assumption was the actual fraction of contaminated sources, which was estimated using a cross-correlation technique with foreground galaxies (Hildebrandt et al., in preparation). If these fractions were biased low, then much of the physical clustering signal would have been interpreted as due to magnification, leading to mass estimates that were too high (at low-$z$).

Currently, we are also investigating the influence of several other systematic effects on our magnification measurements. In particular, we are studying how the varying depth and the varying seeing of the survey affect different lens and source samples, and how stellar contamination (or also just the light haloes of stars) and galactic dust can alter the magnification signal. This in-depth analysis of systematic effects will be presented in a forthcoming paper (Morrison et al., in preparation) and might provide additional insight into the apparent redshift dependence of the cluster magnification signal reported in \citet{Ford14}. Another possibility may be related to the masking effect of cluster galaxy members, which is survey dependent and can affect both magnification and shear measurements \citep{Simet14}. This sky obscuration could lead to our magnification masses being biased low, but our shear measurements should be robust because of the stringent criteria used for selecting background galaxies (Section \ref{sec:measure}).

In order for magnification to yield robust results that encourage its employment in the next generation of large surveys, this discrepancy needs to be addressed. Studies that compare shear and magnification measurements for large binned lens and source samples are crucial for teasing out these underlying systematics. 


\section{Conclusions}

This work has presented weak lensing shear results, measured at 54$\sigma$ significance, for a new catalogue of cluster candidates detected by the 3D-MF algorithm. 3D-MF is a three-dimensional advancement of older matched-filter techniques, which automatically searches wide and deep optical data for galaxy clusters across a range of redshifts. Given a sensible luminosity and radial profile, 3D-MF is able to search within data for a range of galaxy cluster masses. By construction, 3D-MF has allowed us to find lower mass cluster candidates (and groups) which other popular techniques, such as the red sequence and maxBCG, may not be capable of finding. 

3D-MF was run on the CFHTLS-Wide fields using galaxy photometric redshifts and $i'-$band data for cluster luminosity profiles, producing one of the largest and most complete cluster catalogues currently available. 18056 cluster candidates were detected with a significance $\ge 3.5$ and richness $N_{200} >$ 2, out to a redshift of 0.9 ($>$100/deg$^2$). Many of these cluster candidates are in the lower mass ranges (down to $\lesssim 10^{13}\Msun$), which is notably a larger low mass sample than currently exists from deep, wide surveys in the literature, offering an enormous opportunity for further study. 

The CFHTLS-Wide 3D-MF catalogue was investigated to learn more about candidate cluster properties, such as masses and centroiding, as well as to follow up on previous results applying the less developed technique of lensing magnification to this cluster sample \citep{Ford14}. Shear profiles were measured around cluster candidates, which were stacked as a function of richness and redshift, and we focused on presenting composite-halo model fits to measurements of the differential surface mass density $\Delta\Sigma$. 

Careful consideration of potential miscentring of galaxy clusters by 3D-MF had to be taken into account in the analysis. We fit the data with smoothed shear profiles, $\Delta\Sigma^{\rm sm}$, that describe a cluster whose halo is offset from its assumed centre. The fraction of clusters that are affected by miscentring, as well as the probability distribution of the offsets, were both allowed to vary in the modelling. We found the inclusion of these parameters to significantly improve the $\chi^2$ of the cluster profile fits, relative to a perfectly centred model for $\Delta\Sigma$, which we also demonstrated for comparison. The stacked cluster shear measurements were well fitted by a model in which about half the clusters are affected by miscentring ($p_{\mathrm{cc}} \sim 0.5$), with the distribution of centroid offsets peaking at $\sim$ 0.4 arcmin.

The large sample of cluster candidates in this work allowed us to bin the shear measurements as a function of both richness and redshift. The average cluster candidate masses were found to be relatively constant with redshift, estimated at 2 to 4 $\times 10^{13} M_{\odot}$. The masses scaled strongly with richness, ranging from $\sim 6 \times 10^{12} M_{\odot}$ to $\sim 3 \times 10^{14} M_{\odot}$. We measured the normalization and slope of the mass-richness relation for the 3D-MF cluster candidates, finding $M_0$ = $(2.7^{+0.5}_{-0.4}) \times 10^{13} M_{\odot}$ and $\beta$ = $1.4 \pm 0.1$. The redshift dependence of the normalization $M_0 (z)$ was not significant, yielding a powerlaw slope in $(1+z)$ of $-0.4^{+0.5}_{-0.6}$. 

The masses of individual cluster candidates were found to range from a small group scale, with stacked average masses of less than $10^{13} M_{\odot}$, all the way up to a few very massive clusters, at several $10^{15} M_{\odot}$. Since the 3D-MF catalogue has not been followed up spectroscopically, we expect some fraction of false-detections (estimated between $\sim 16$ and $24$\% from simulations), which would lead to these mass estimates being biased low, and would especially affect the low-richness stacked measurements. We note, however, that the impact of false detections may be less severe than implied, if line-of-sight projections are significant. Chance alignments of low-mass structures would have a similar effect on a shear measurement (which probes surface mass density) as it would on an estimate of optical cluster properties like richness.

By design, we binned cluster candidates in an identical fashion to the previous magnification study \citep{Ford14}, and compared the results obtained. This is the first large study directly comparing the outcomes of magnification and shear on the same lens sample. When stacked across all redshifts, we found that the average masses derived within a given richness bin were similar (within 1$\sigma$), but magnification masses were systematically lower, yielding a 2$\sigma$ difference in the normalization of the mass-richness relation derived from the two techniques. The mass-richness slope was essentially identical for magnification and for shear. The comparison across redshift slices yielded very interesting insights into problems that may still exist for magnification. The fact that the shear-determined masses were roughly constant across redshift led us to conclude that the magnification measurement (using magnification-biased number counts of Lyman-break galaxy sources) may still suffer from residual systematics at low-$z$. Notably, however, this occurs at very predictable lens redshifts, so if one has accurate photometric redshift distributions for the sources, these contaminated redshift zones could potentially be avoided.

In future work, it would be interesting to apply various cluster finding algorithms to the {\em same} large data set in order to compare the capabilities of the finders, potentially increase the overall cluster sample and reduce its biases, or even just to compare how different search algorithms perform.  This could ideally lead to more complete and unbiased cluster samples. Current surveys, such as the Kilo-Degree Survey \citep{KiDS13}, the Subaru Hyper Suprime-Cam Project \citep{HSC10}, and the Dark Energy Survey \citep{DES05} for example, are large enough that cluster masses and concentrations will be measured quite accurately as a function of redshift and richness. The area of these surveys is an order of magnitude higher than the CFHTLS-Wide, and high precision cluster profiling will naturally continue to evolve alongside these surveys.

The CFHTLS-Wide 3D-MF galaxy cluster catalogue contains 18056 cluster candidates, over a wide range of mass and redshift, and is now publicly available at \url{cfhtlens.org}. We encourage others to make use of the rich science opportunities afforded by this catalogue. 


\section*{Acknowledgements}
The authors would like to thank Jasper Wall for helpful discussions on statistics, and the referee for providing helpful feedback and suggestions that greatly improved this work. JF is supported by a UBC Four-Year-Fellowship and NSERC. LVW is supported by NSERC and CIfAR. HHi is supported by the DFG Emmy Noether grant Hi 1495/2-1. TE is supported by the Deutsche Forschungsgemeinschaft through project ER 327/3-1 and the Transregional Collaborative Research Centre TR 33 - ``The Dark Universe''. CH, AC, and NR acknowledge funding from the European Research Council under the EC FP7 grant number 240185. LF acknowledges support from NSFC grants 11103012 \& 11333001, Innovation Programme 12ZZ134 of SMEC, STCSM grant 11290706600, Pujiang Programme 12PJ1406700 and Shanghai Research grant 13JC1404400. MH is supported by NSERC.

This work is partly based on observations obtained with MegaPrime/MegaCam, a joint project of CFHT and CEA/IRFU, at the CFHT which is operated by the NRC of Canada, the Institut National des Sciences de l’Univers of the Centre National de la Recherche Scientifique (CNRS) of France, and the University of Hawaii. This research used the facilities of the Canadian Astronomy Data Centre operated by the National Research Council of Canada with the support of the Canadian Space Agency. CFHTLenS data processing was made possible thanks to significant computing support from the NSERC Research Tools and Instruments grant programme.

{\it Author Contributions:} All authors contributed to the development and writing of this paper. The authorship list reflects the lead authors of this paper (JF, LVW, MM, CL and HHi) followed by two alphabetical groups. The first alphabetical group includes key contributors to the science analysis and interpretation in this paper, the founding core team and those whose long-term significant effort produced the final CFHTLenS data product. The second group covers members of the CFHTLenS team who made a significant contribution to the project and/or this paper. The CFHTLenS collaboration was co-led by CH and LVW.


\bibliography{References}

\end{document}